\documentstyle[11pt,epsfig]{article}
\title{Contribution of Orbital Angular Momentum to the Nucleon Spin
}

\author{ Maryam Momeni-Feili$^{(1)}$, Firooz Arash $^{(2)}$ \footnote{Corresponding author:farash@cic.aut.ac.ir },
 Fatemeh Taghavi-Shahri$^{(3)}$, Abolfazl Shahveh$^{(4)}$
\\
$^{(1)}$ Department of Physics, Science and Research Branch,Islamic Azad University, Tehran, Iran\\
$^{(2)}$ Physics Department, Tafresh University, Tafresh, Iran \\
$^{(3)}$ Department of Physics, Ferdowsi university of Mashhad, P.O. Box 1436, Mashhad, Iran \\
$^{(4)}$ Diamond Light Source Ltd., Diamond house, Herwell Science and innovation Campus \\ Didcot, Oxfordshire OX11 0DE, Uk 
 }

\begin{document}
\maketitle

\begin{abstract}
We have calculated the Orbital Angular Momentum of quarks and gluons in the nucleon.
The calculations are carried out in the next to leading order utilizing the
so-called valon model. It is found that the average quark orbital
angular momentum is positive, but small, and the average gluon
orbital angular momentum is negative and large. We also report on some regularities about the
total angular momentum of the quarks and the gluon, as well as on the orbital angular momentum of the
separate partons. We have also provided partonic angular momentum, $L^{q,g}$ as a function of $Q^2$.
\end{abstract}

\section{INTRODUCTION}
Polarized deep inelastic scattering processes is the most direct
tool to probe the spin content of the nucleon. In such
experiments detailed information can be extracted on the shape and
the magnitude of the spin dependent parton distributions, $\Delta
q_{f}(x,Q^{2})$. Deep inelastic scattering reveals that the
nucleon is a rather complicated object consisting of an infinite
number of quarks, anti-quarks, and gluons. It is a common belief
that other strongly interacting particles also exhibit similar
internal structure. \\
The decomposition of nucleon spin in terms of its constituents has
been a challenging and an active topic in hadron physics, both from
theoretical and experimental points of view. It is now established
that quarks carry a small fraction of
the nucleon spin. Other sources that might contribute to the
nucleon spin are gluon spin and the overall orbital
angular momentum of the partons. Thus, it is common to write the
following spin sum rule for a nucleon.
\begin{equation}
\frac{1}{2}=\frac{1}{2}\Delta \Sigma +\Delta G +L_{q,g}
\end{equation}
Over the past few years we have studied the first two terms of the
above sum rule within the framework of the so called {\it{valon}}
model \cite{arash1} in the next to leading order. The valon model is a
phenomenological model for hadrons, introduced first by R.C. Hwa
about thirty years ago \cite{hwa1}. The model has been quite
successful in describing a variety of hadronic phenomena \cite{hwa2,arash11}.
In the polarized deep inelastic scattering domain, the model has
successfully reproduced the existing data on $g_{1}^{p,n,d}$, and
individual parton contributions, $\Delta q_{f}(x,Q^{2})$, to the
nucleon spin, while predicting new results yet to be tested. Among other things, the
model has predicted that the sea quark polarization is negligible,
which is now proven to be the case\cite{Airapetian}. This finding is because the valons are generated by perturbative dressing in QCD.
In such processes with massless quarks, helicity is conserved and
therefore, the hard gluons can not produced the sea quark polarization
perturbatively. So, it turns out that
sea polarization is consistent with zero.
 We have also shown that
although $\delta g(x,Q^2)$ is small, but its first moment, $\Delta G$ is large
and grows as $Q^2$ increases \cite{Shahveh}. This is consistent with
QCD and with available experimental data \cite{data1},\cite{data2},\cite{data3},\cite{data4},
\cite{data5},\cite{data6},\cite{data7},\cite{data8},\cite{data9} .
Elsewhere, we have reported that with a fixed and almost $Q^2$ independent
value for $\Delta \Sigma$ and with the growing $\Delta G$ it is not
possible to achieve $s_{z}=\frac{1}{2}$ for the nucleon spin. Therefore,
there should be other contributing element in order to arrive at
spin $\frac{1}{2}$ of the nucleon. The only possible place would be
the orbital angular momentum of the partons. In \cite{arash2} we used Eq.
(1) and estimated, but not independently calculated, the magnitude
of the overall orbital angular momentum of the partons inside the nucleon.
Our conclusion was that overall orbital angular momentum of
partons is negative and decreases as $Q^2$ increases. In
fact, almost twenty five years ago, P. G. Ratcliffe \cite{ratcliffe}
suggested that a consistent interpretation of the
Dokshitze-Gribov-Altarelli-Parisi evolution equation of helicity
weighted parton distributions requires the partons to carry a sizable orbital angular
momentum. Moreover, he concluded that for large $Q^2$,
the average orbital angular momentum will be negative. \\
The purpose of this paper is to report the results that we have
obtained for the orbital angular momentum contribution of quarks
and gluons to the proton spin. Our calculations are carried out in
the next to leading order in perturbation theory and are based on the
valon representation of the nucleon. It is important to mention that
one could, as well, use other polarized and unpolarized parton
distributions, such as those obtained from the available global
fits to carry out the same calculation.  The main reason for us in using the valon
model is that, first, it was handy and secondly, it's outcomes
and the predictions have proven to be consistent with all the
experimental data that are available. Hence, providing a reasonable
confidence in its physical validity.  \\

\section{The Experimental data}

Over the past two decades theoretical framework for the
understanding of the spin structure of nucleon has been developed
and numerous experiments were carried out. We now have a fairly good
understanding of the first term in the sum rule above. Total quark
contribution to the nucleon spin is about $\Delta \Sigma =0.4$. Some data are also
emerged on the gluon polarization \cite{Stolarski}\cite{Djawatho} showing that
$\frac{\delta g(x,Q^2)}{g(x,Q^2)}$ is small. These data are
reported at individual kinematics, i.e. at separate $(x, Q^2)$
points, and lack the same level of precision achieved for the
quark sector. The  data from COMPASS collaboration \cite{Alekseev}
may even be able to rule out a negative value for $\frac{\delta
g(x,Q^2)}{g(x,Q^2)}$, which has been a controversy over the past
few years. Nevertheless, the smallness of $\frac{\delta g}{g}$ by
itself cannot rule out the possibility of a large value for the
first moment, $\Delta G$, of the gluon polarization. In Figure 1
we present $\frac{\delta g(x,Q^2)}{g(x,Q^2)}$ that is obtained from
our model. We have calculated $\frac{\delta g(x,Q^2)}{g(x,Q^2)}$ at
each kinematical point for which the data exists. The apparent wide
band in the figure is actually several closely packed curves
corresponding to the several values of $Q^2$ at which data points
are measured. The details of this calculation can be found in
\cite{Shahveh}. In reference \cite{Binder} a new method is suggested for measuring
the gluon polarization.
In Figure 2 we also show our results on the first moments of the
quark, $\Delta \Sigma$, and the gluon, $\Delta G$, polarization in
the nucleon. Substituting these valued in Equation 1, gives a
measure of the total overall angular momentum of the partons inside
the nucleon, which is also shown in figure 2.

\begin{figure}[htp]
\centerline{\includegraphics[width=14cm]{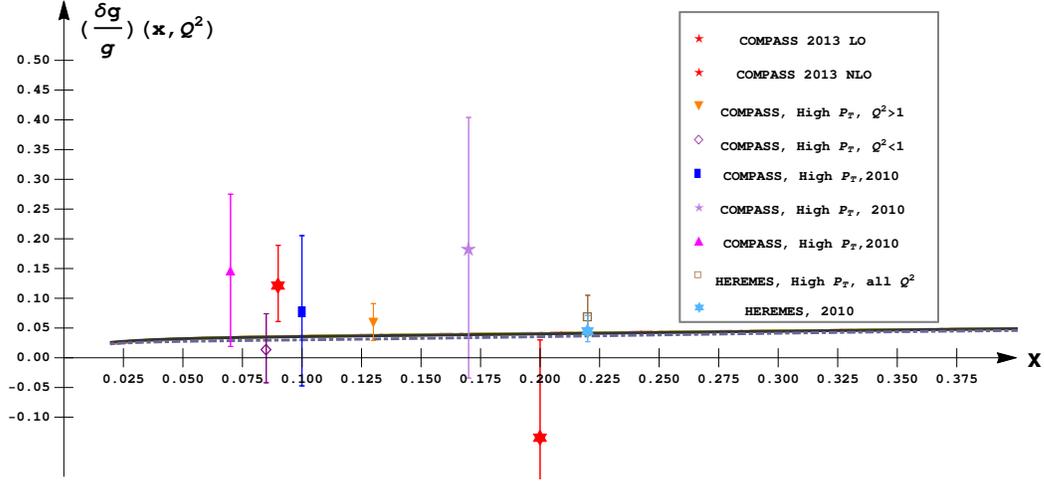}}
\caption{The ratio
$\frac{\delta g}{g}$ calculated in the valon model and compared with
the data. Data point are from \cite{data1},\cite{data2},\cite{data3},\cite{data4},\cite{data5},\cite{data6},\cite{data7},\cite{data8}, \cite{data9} } \label{figure 1.}
\end{figure}

\begin{figure}[htp]
\centerline{\includegraphics[width=10cm]{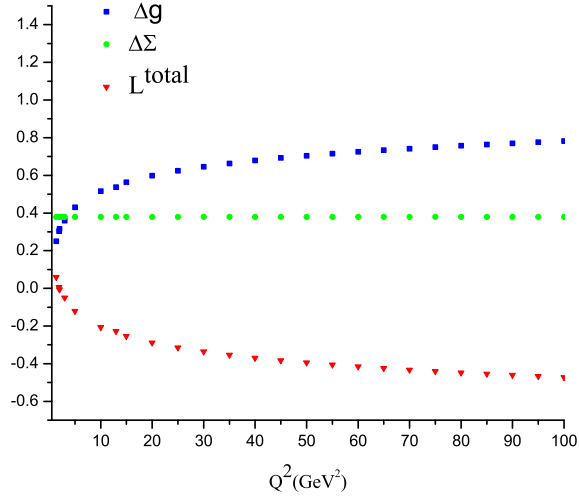}}
\caption{First
moments of polarized gluons and quarks distribution functions and the resulting total
orbital angular momentum obtained from Equation.1 } \label{figure
2.}
\end{figure}

\section{Orbital Angular Momentum}
Considering that the spin of quarks accounts for only a part of the nucleon
spin, and that of the gluon is still unclear, a substantial fraction of the
nucleon spin must be due to the orbital angular momentum.
Unfortunately, in the gauge theories there is no unique decomposition of the
nucleon spin into contributions due to spin and the orbital angular momentum
of quarks and gluons. For example, Jaffe and Manohar \cite{Jaffe} have used a
light-like hypersurface, employed the light-cone
framework and light-cone gauge and arrived at the following decomposition
\begin{equation}
\frac{1}{2} = \frac{1}{2}\sum_q \Delta q +\sum_q \mathcal{L}^{z}_{q} + \Delta G + \mathcal{L}^{z}_{g}
\end{equation}
In this decomposition each term is defined as the matrix element of the
corresponding term in $+12$ component of the orbital angular momentum tensor.
The first and the third terms are interpreted as the quark and
gluon spin, respectively. The second and the forth terms are identified as the
quark and gluon orbital angular momentum. In this decomposition, except for the
first term, individual terms are not separately gauge invariant.
An alternative decomposition is provided by Ji \cite{Ji1}
\begin{equation}
\frac{1}{2} = \frac{1}{2}\sum_q \Delta q +\sum_q L^{z}_{q}+ J_{g}^{z}
\end{equation}
where each term is separately gauge invariant. However, the gluon total angular momentum
is not decomposed, in a gauge invariant way, into its spin and the orbital angular momentum.
In Ji's decomposition, The total spin of quarks, $J_q$, and that of the gluons, $J_g$, are related
to the generalized parton distribution (GPD) at twist-two level.
Other decompositions have also been proposed \cite{Chen}\cite{Wakamatsue}. A thorough analysis of these
decompositions is given in \cite{Leader}. Briefly, it has been established that there are only
two types of complete decompositions of the nucleon spin. The first one is the
decomposition of canonical type, while the other is the decomposition of mechanical (or kinetic) type.
 The famous Jaffe-Manohar decomposition belong
to the former, while another complete decomposition proposed in \cite{Wakamatsue} is
of the mechanical type. Since these two
quark orbital angular momenta (OAMs) are apparently different, the gluon
OAMs are also different in the two types of nucleon spin decomposition.\\
It is now shown that at the twist-three level, once the generalized parton
distribution are integrated over $x$ , both decompositions given in eq. (2) and Eq. (3)
can be obtained. \cite{Ji2}, \cite{Hatta}.\\
As mentioned, Ji's decomposition are related to twist-two generalized parton distributions,
which can be measured in deeply virtual compton scattering. The quark orbital angular
momentum in terms of GPDs is given by \cite{Ji3}
\begin{equation}
L^q = \int dx \int d^2b (x H^q(x,b)+x E^q(x,b)- \tilde H^q(x,b))
\end{equation}
The GPDs describe the dynamics of partons in the transverse plane
in position space. Complementary information on the dynamics of partons
in the transverse plane, but in the momentum space, is obtained from
Transverse Momentum Dependent parton distributions (TMD-PDF)\cite{Boer} \cite{Anselmino}.
Therefore, one naturally expects that TMD and GPDs will teach us about partonic
orbital angular momentum.\\
The orbital angular momentum of partons play an important role in hadron physics.
It is well known that in order to have a non-zero anomalous magnetic moment,
the light cone wave function of nucleon must have components with
$L_z \neq 0$ \cite{Burkardt}.\\
In the local limit, GPDs reduce to form factors, which are obtained from
the matrix elements of the energy momentum tensor $\Theta ^{\mu \nu}$. Since
one can define $\Theta ^{\mu \nu}$ for each parton, one can identify the momentum
fraction and contributions to the orbital angular momentum of each
quark flavor and gluon in a hadron. Spin flip form factor $B(q^2)$
which is the analog of the Pauli form factor $F_2(Q^2)$ of the nucleon
provides a measure of the orbital angular momentum carried by each
quark and gluon constituent of the nucleon at $q^2=0$. Similarly,
the spin conserving form factor $A(q^2)$, the analog of Dirac form
factor $F_1(q^2)$, allows one to measure the momentum fraction carried
by each constituent. This is the underlying physics of Ji's sum rule \cite{Ji3}:
\begin{equation}
J^z_{q,g}=\frac{1}{2}[A_{q,g}(0)+B_{q,g}(0)]
\end{equation}
where, $B_{q,g}$ are the second moments of unpolarized spin-flip GPD in the
forward limit. It is subject to the constraint that
\begin{equation}
B(0)=\sum_i B_i(0)=0
\end{equation}
that is, when summed over all partons, spin flip form factor vanishes.
For composite systems, this has been proven by Brodsky, {\it{et al.}} in the
light cone representation\cite{Brodsky}. In fact, it is a
consequence of equivalence principle. \\
For the quark and the gluon sectors, the above equation translates into
\begin{equation}
J^z_q (x)=\frac{1}{2}x[<q(x)>+B_q(0)],        J^z_g (x)=\frac{1}{2}x[<g(x)>+B_g(0)].
\end{equation}
Based on some lattice calculations and the model dependent
analysis \cite{Wakamatsu2} it is expected that $B_{q,g}$ to be small.
In fact, lattice calculations show that the valence quarks give a value
between $-0.077$ and $0.015$ \cite{Bratt}.
Excluding the unlikely possibility for large value due to strange quark
and anti-quark, we find that the sum of the contributions from the sea quarks and the
gluons must be small. We do realize the possibility that gluon and the sea quark
contributions could be large, but with opposite sign. Also, they can be
large, but have nodes such that their second moments become small. Lattice
calculations \cite{Liu} have verified that indeed the total anomalous gravitomagnetic
moment of the nucleon is zero. A more Recent lattice
calculation \cite{Deka} have shown that $B_{q,g}=0.00(6)$.
They have used a different notation, namely $T_{2}(0)$, and presented their results in Table III of their paper. \\
\begin{table}
 {\footnotesize
\centerline{\begin{tabular}{|c|c|c|c|}
  \hline
  $Q^2$ & $1.9 GeV^2$ & $5 GeV^2$ & $10 GeV^2$ \\
  \hline
   \hline
$\Delta \Sigma$ & 0.38 & 0.38 & 0.38 \\
 \hline
$\Delta g$ & 0.303 & 0.440 & 0.516 \\
 \hline
$L_q$ & 0.0944 & 0.0723 & 0.0616 \\
 \hline
$L_g$ & -0.0848 & -0.275& -0.310\\
   \hline
\end{tabular}}
 \caption{The numerical results for $\Delta \Sigma, \Delta g, L_q, L_g$ in the Valon model.}}
\end{table}
Therefore, in the following analysis we will set $B_{q,g}=0$. With a zero value
for $B_{q,g}$, the "mechanical orbital angular momentum" of partons, $L_{q,g}$, can be determined
entirely from polarized and unpolarized parton distributions. Moreover, with such an assumption
the evolution equation for the angular momentum distributions $J_{q,g}$
is exactly the same as that for the unpolarized quark and gluon distributions \cite{Hoodbhoy}. These
distributions are evaluated by us in the valon model with good accuracy in a wide range of kinematics
$Q^2 =[0.4, 10^6]$ $GeV^2$ and $x=[10^{-6}, 0.95]$ and will be utilized here. The details and
the functional form of the unpolarized parton distributions in the valon representation can be found in \cite{arash11}.
The table .1 show the numerical results for $\Delta \Sigma, \Delta g, L_q, L_g$ in some values of $Q^2$.\\
In Figure 3 we show the behavior of $L_q(Q^2)$, and $L_g(Q^2)$ at several $Q^2$ values.
It is apparent that while the quark orbital angular momentum is small and positive, the gluon
orbital angular momentum is negative and decreases as $Q^2$ increases.
\begin{figure}[htp]
\centerline{\includegraphics[width=10cm]{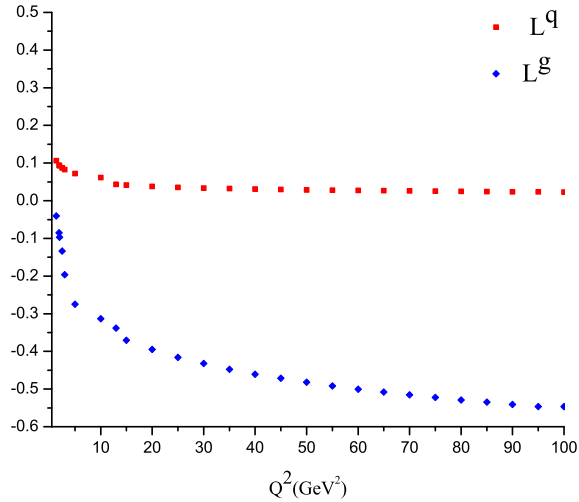}}
\caption{Orbital angular
momentum of quarks and gluons in the valon model as a function of $Q^2$} \label{figure 3.}
\end{figure}
We have checked to make sure that if our results reproduces $J^{p}=\frac{1}{2}=J_{q}+J_{g}$.
The results are shown in Figure 4. Evidently, this is the case.
\begin{figure}[htp]
\centerline{\includegraphics[width=10cm]{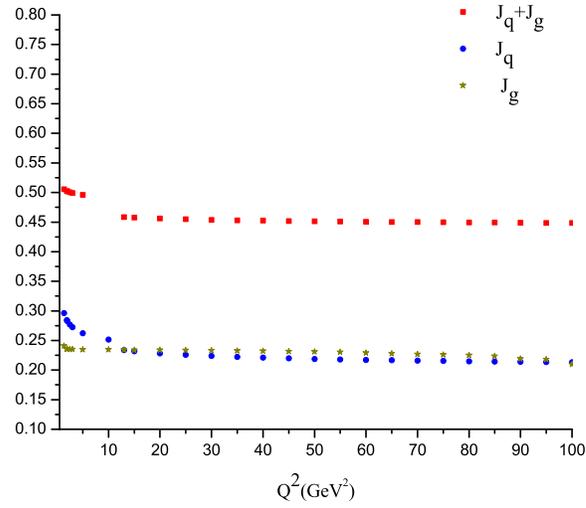}}
\caption{Total angular
momentum of quarks and gluons in the valon model} \label{figure 4.}
\end{figure}
In Figure 5 we present the gluon spin, $\Delta g$, the gluon orbital angular momentum, $L_{g}$
and the total angular momentum as a function of $Q^2$. This figure indicates that $J_{g}$
is independent of $Q^2$ and contributes an amount of about $0.22$ to the nucleon spin.
\begin{figure}[htp]
\centerline{\includegraphics[width=10cm]{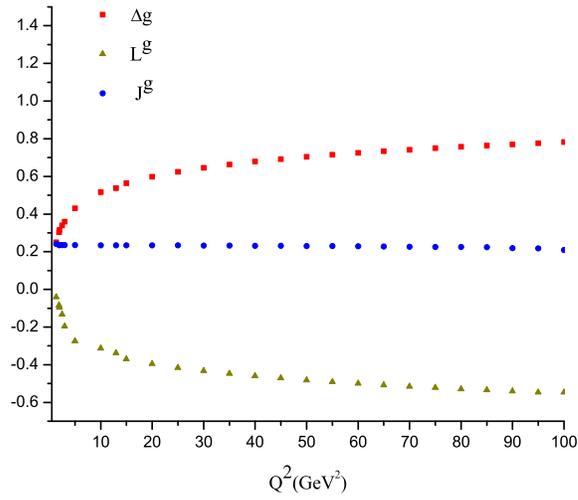}}
\caption{Spin, orbital angular
momentum and the total angular momentum of gluons in the valon model} \label{figure5.}
\end{figure}
in Figure 6 the total angular momentum of individual quark flavors, $J_{u}$ and $J_{d}$,
are presented. In this figure we also compare  our results for $J_{u,d}$  with
those of Ref. \cite{Bratt,Taneja1,Bacchetta,Guidal,Göckeler,Goloskokov,Ohtani,Hagler}. We further found that the orbital angular momentum of u-quark, $L_{u}$, and
d-quark, $L_{d}$, have opposite signs and largely cancel each other.
Our results indicate that $L_{d}$ is positive and $L_{u}$ is negative.
This is shown in Figure 7 in comparison with those from \cite{Bratt,Taneja1} . Their difference is shown in Figure 8 and
seems that the dependence on $Q^2$ is marginal. In an interesting paper \cite{Taneja1} authors have
derived a sum rule for spin-1 system through which they have obtained the total and the
orbital angular momenta for $u$ and $d$ quarks in the proton at $Q^2=4$ $GeV^2$. Our results on total
angular momentum carried by quarks, that is, $J_{q}=J_{u}+J_{d}$ agrees with
the findings of Ref. \cite{Taneja1}, amounting to 0.26 at $Q^2=4$ $GeV^2$.  This is interesting, because the two
approaches are quite different. The two approaches also agree on the sign of $L_{u}$ and
within the errors, the numerical values are also compatible. Our findings, however, is different
from those of \cite{Taneja1} on the total and the orbital angular momenta of the $d$ quark. Yet,
both approaches produce compatible values for the spin component of the the $d$ quark.
The total quark orbital angular momentum, $L^{Q}=L_{u}+L_{d}$ in our model gives a value of $0.08$ at
$Q^2=4$ $GeV^2$, whereas, the result of the Ref. \cite{Bratt} is $-0.016\pm 0.084$. An earlier estimate of
$L^{Q}=0.05 - 0.15$ is also given by Ji and Tang \cite{Ji3}. It is evident that within the quoted errors,
the two values are not far apart. In fact, except for $L_{d}$, within the errors, our
results are fairly close to those of Ref. \cite{Bratt}. We find
no crossover between $L_u$ and $L_d$ when $Q^2$ is varied. We also find  that $L_u - L_d$ remains  large and negative
 and this finding is in nice agreement with \cite{wakamatsu3}. \\
\begin{figure}[htp]
\centerline{\includegraphics[width=10cm]{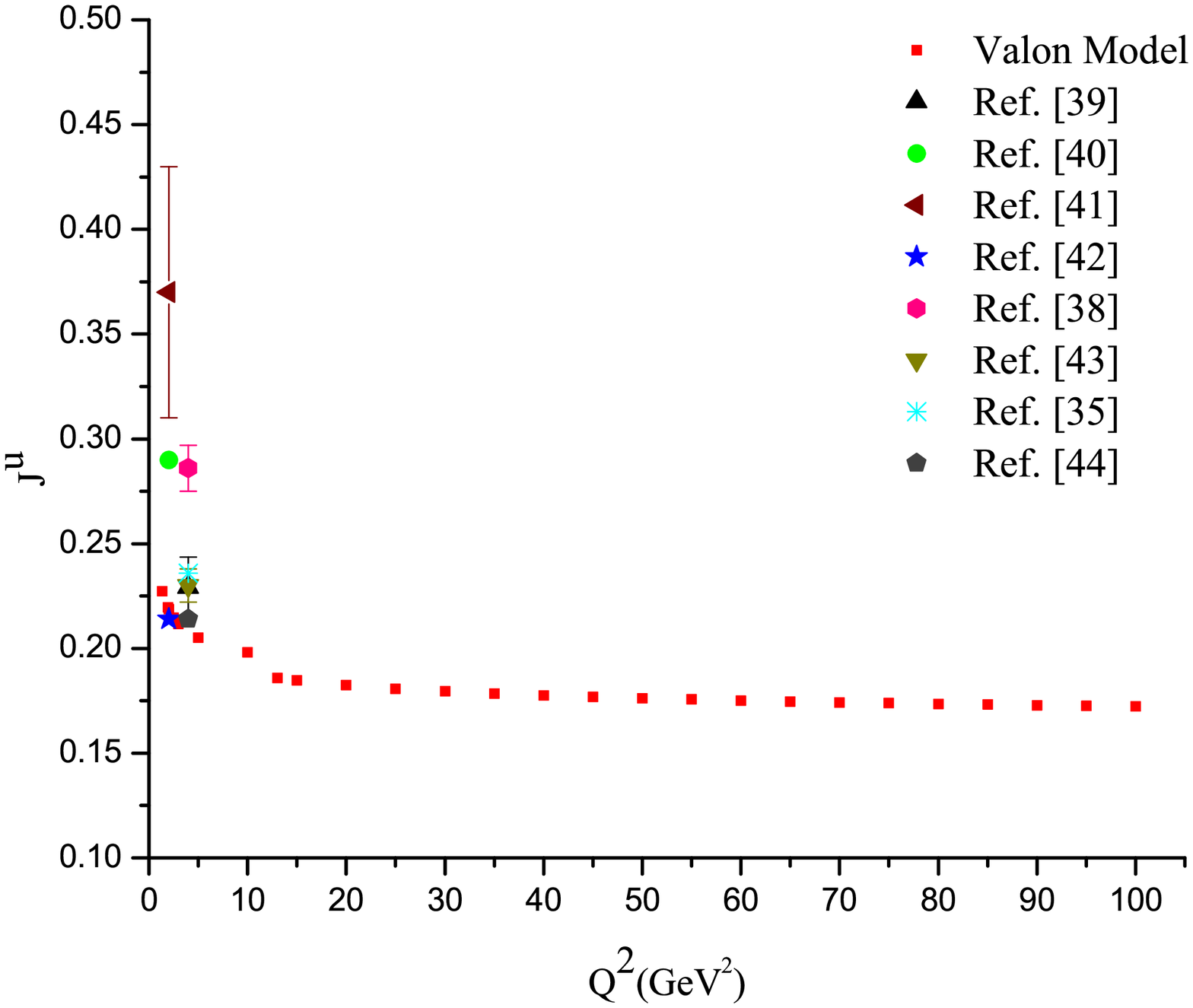}}
\centerline{\includegraphics[width=10cm]{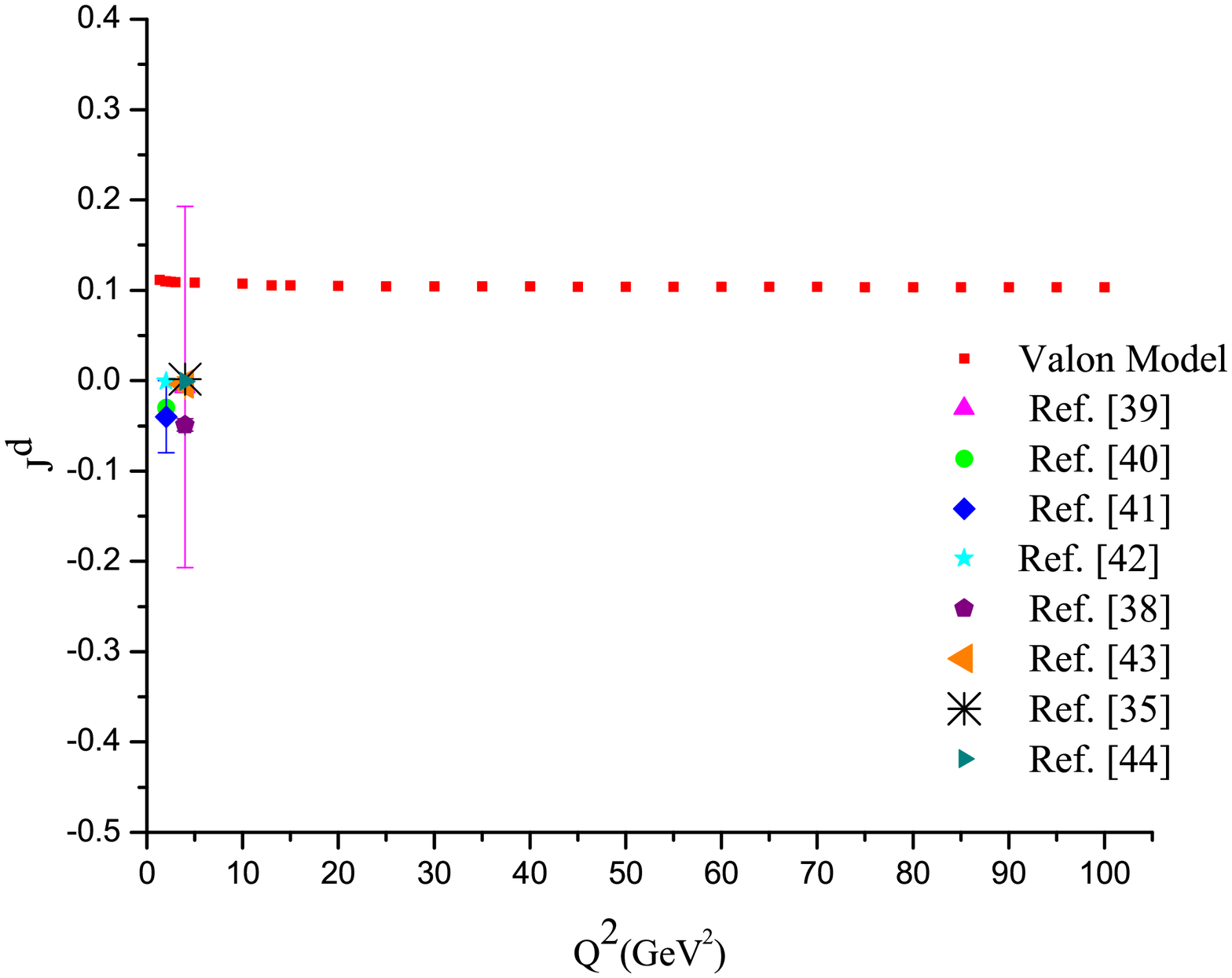}}
\caption{Total angular
momentum for u-quark and for d-quark and comparison with other models.} \label{figure6.}
\end{figure}
\begin{figure}[htp]
\centerline{\includegraphics[width=10cm]{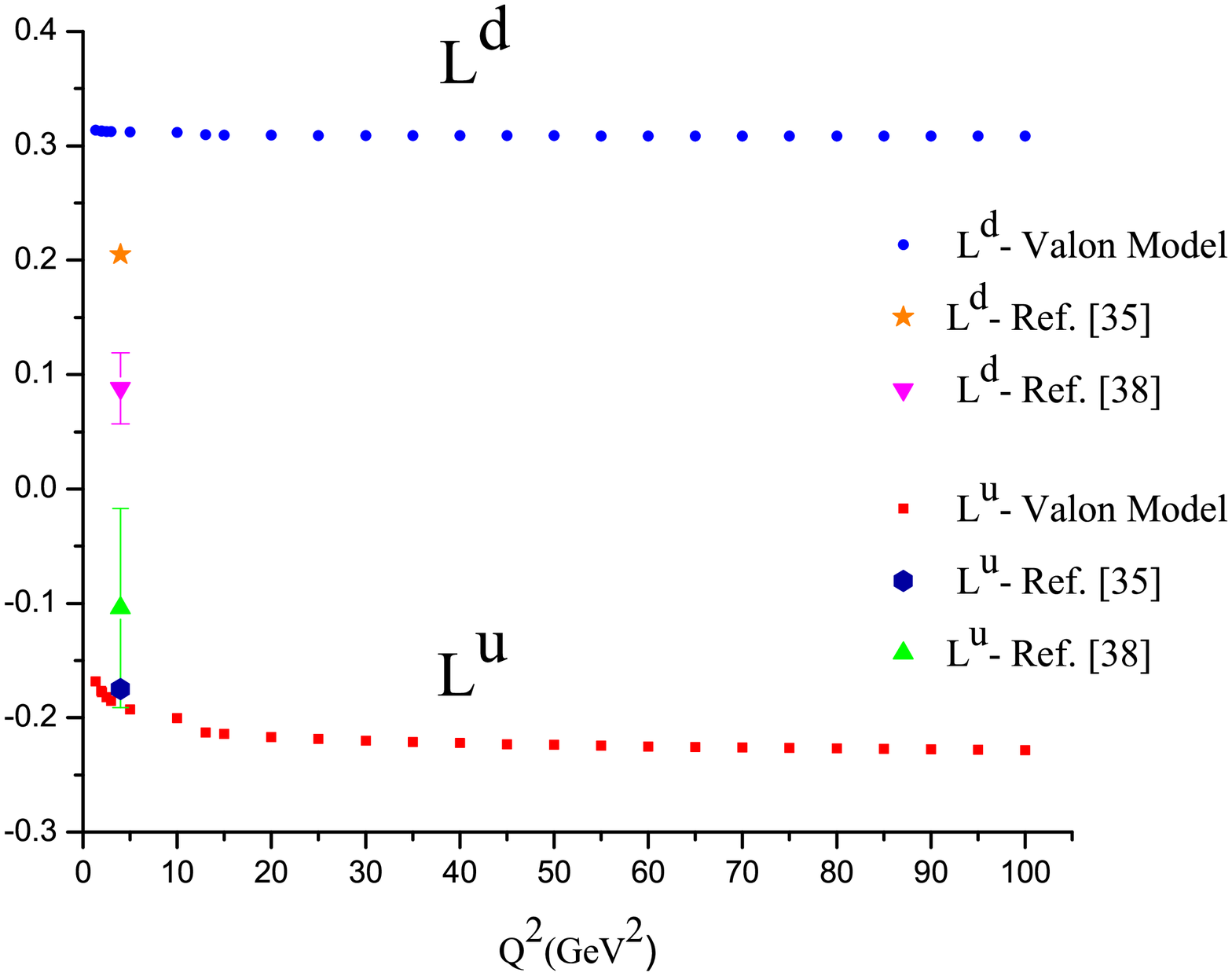}}
\caption{Orbital angular
momentum for u-quark and for d-quark and comparison with other models.} \label{figure7.}
\end{figure}
Finally, we note that $Q^2$ dependence of $J_{g}$ is marginal, and the interplay is
between $\Delta g$ and $L_{g}$, the former increases with $Q^2$, while the latter decreases.
This is evident from Figure 5. It is also interesting to mention that above $Q^2$
around $5$ $GeV^2$ or so, the total angular momentum of quarks and the gluons seems
to approach to an identical value, indicating that they equally share the spin of the nucleon.
This observation is manifestly apparent from Figure 4.
\begin{figure}[htp]
\centerline{\includegraphics[width=10cm]{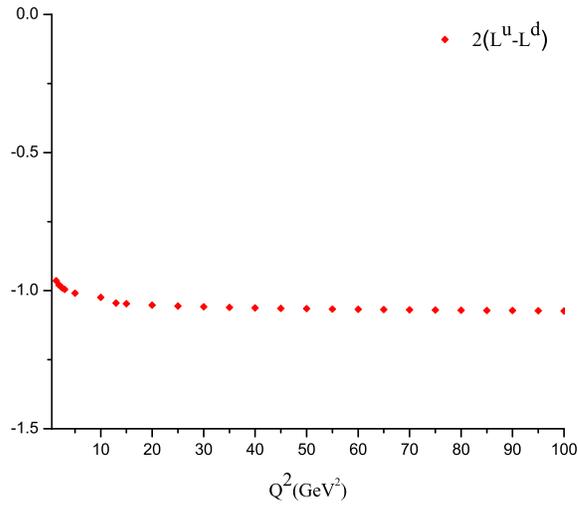}}
\caption{The difference
between orbital angular momentum of the u-quark and the d-quark.} \label{figure7.}
\end{figure}

\section{conclusion}
Although the valon representation of hadrons is a simple phenomenological model, it describes the
structure of nucleon rather nicely. within this model, we have investigated the orbital angular momentum
contribution of quarks and gluons to the nucleon spin. It shows that the quarks orbital angular momentum
contribution to the total angular momentum of the nucleon is positive and relatively small. However,
the gluon orbital angular momentum contributes substantially. Thus, we conclude that Gluon is a major
player in describing the spin structure of nucleon. On the one hand, while $\frac{\delta g}{g}$ is small, first
moment of the gluon polarization, $\Delta G$, is large and increases as $Q^2$ grows. On the other hand, its
orbital angular momentum is large and negative, thus compensating the growth of $\Delta G$.\\
Some regularities also have emerged from our study: both orbital and total angular momenta of
the u-quark and the d-quark seems to be independent of $Q^2$, though, some $Q^2$ dependence for $J_{u}$ is
observed at low $Q^2$, but rapidly disappears. The same is true for the total angular
momentum of the gluon, However, its orbital angular momentum varies.
These are evident from figures 7 and 5, respectively.
Finally, we notice that our calculations seems to be compatible with those that are obtained in \cite{Bratt}.
We have also presented various orbital angular momenta components as a function of $Q^2$ which may be utilized to gain information on some generalized parton distributions.

\section{Appendix: The Valon Model}
Our understanding of hadron structure comes from the deep inelastic data and the hadron spectroscopy. In the latter picture, hadrons are bound states of massive particles, loosely called "quarks" or "constituent quarks". The bound states of those entities describe the static properties of the hadrons. On the other hand, the interpretation of the deep inelastic data relies on the quarks of the QCD Lagrangian with a very small mass. The hadronic structure in this picture is intimately connected with the presence of a large number of partons (quarks and gluons). The quarks that participate in the bound state problem and the quarks of the QCD Lagrangian differ in other important properties as well. The very obvious example is the color charge of quark field in QCD Lagrangian, which is not gauge invariant and, thus, ill defined; reflecting the color of gluons in an interacting theory. Whereas, color associated with the quarks of a bound state (constituent quark) is a well defined entity.\\
In the bound state problem we regard a proton as consisting of three quarks and pion, a quark-antiquark pair. These are the constituent quarks. In deep inelastic scattering a proton is viewed as having valence quarks, sea quarks, and gluons, collectively called partons. In fact, even at $Q^2$ as low as a few $GeV^2$ the gluons carry nearly half of the nucleon momentum. To reconcile the two pictures of hadron, it is necessary to realize that the quarks probed in deep inelastic scattering are current quarks of the QCD Lagrangian and not the constituent quarks of the bound state problem. The failure to recognize this difference can lead to many mistakes.\\
By definition, a valon is a structureful object consisting of a valence quark plus its associated cloud of sea quarks and gluons. The cloud, or the structure of the valon is due to the dressing process in QCD. Indeed, it is shown \cite{Lavelle1}\cite{Lavelle2} that one can dress a QCD Lagrangian field to all orders in perturbation theory and construct such an object (which we called a {\it{valon}}) in conformity with the color confinement. From this point of view, a valon emerges from the dressing of a valence quark with gluons and $q\bar{q}$ pair in QCD. In a bound state problem those processes are virtual and a good approximation for the problem is to regard a valon as an integral unit whose internal structure cannot be resolved. Thus, it may be identified as an indivisible , point-like object. As such, in a bound state problem they interact with each other in a way that is characterized by the valon wave function. On the other hand, in a scattering process, the virtual partons in a valon can be excited and put on mass-shell. They respond independently in an inclusive hard collision with a $Q^2$ dependence that can be calculated in QCD at high  $Q^2$. The point is that the valons play a dual role in hadrons: on the one hand, they are constituents of bound state problem involving the confinement at large distances. On the other hand, they are quasi-particles whose internal structure are probed with high resolution and are related to the short distance problem of current operators. This picture suggests that the structure function of a hadron involves a convolution of two distributions, namely, the valon distribution in the hosting hadron and the parton distribution in the valon. In an unpolarized situation on may write the structure function of hadron $h$ as follows
\begin{equation}
F_{2}^{h}(x,Q^2)=\sum_{valon}\int_{x}^{1}dy G_{valon}^{h}(y)F_{2}^{valon}(\frac{x}{y},Q^2)
\end{equation}
 where $F_{2}^{valon}(\frac{x}{y},Q^2)$ is the structure function of the probed valon and can be calculated in perturbative QCD. The function$ G_{valon}^{h}(y)$ represents the valon distribution in the hosting hadron carrying momentum fraction $y$ of the hadron.It is $Q^2$ independent. These functions are already calculated for a number of hadrons. Details for the proton can be found in \cite{arash11}.\\
 Similarly, for a polarized hadron, we can write the polarized structure function,$g_{1}^{h}$ as
 \begin{equation}
g_{1}^{h}(x,Q^2)=\sum_{valon}\int_{x}^{1}dy \delta G_{valon}^{h}(y) g_{1}^{valon}(\frac{x}{y},Q^2)
\end{equation}
where, $\delta G_{valon}^{h}(y)$ is the helicity distribution of valon in the hadron with momentum fraction $y$ of the hadron. $g_{1}^{valon}(\frac{x}{y},Q^2)$ is the polarized structure function of the valon. Again, the detailed calculation of these functions are given in \cite{arash1}.  \\ We have worked in
$\overline{MS}$ scheme with $\Lambda_{QCD}=0.22$ $GeV$. The initial scale of energy is
$Q_{0}^{2}=0.283$ $GeV^2$. The  motivation for this initial inputs
at $Q_{0}^{2}$ comes from the phenomenological consideration that
requires us  to choose the initial input densities as $\delta(z -
1) $ at $Q_{0}^{2}$. This condition means that the internal structure of the valon
can not be resolved at $Q_{0}^{2}$  and  at this initial scale, the nucleon can
be considered as a bound state of three valence quarks that carry
all the momentum and the spin of the nucleon. Increasing the  $Q^2$ values resolved the other partons in the nucleon. Therefore,  our initial input densities to solve the DGLAP
equations inside the valon are

\begin{eqnarray}
\delta q^{NS }(z,Q_0^{2})= \delta q^{S }(z,Q_0^{2})= \delta(z -
1)\\
\delta g(z,Q_0^{2})=0
\end{eqnarray}

In the valon picture of hadron, the deep inelastic scattering with high enough $Q^2$ actually it is the structure of a valon that is probed. At low $Q^2$ the valon structure cannot be probed and hence behaves as a quark in the bound state problem. This means that if $Q^2$ of the probe is less than a threshold value of $Q_0$ then, a valon would appear as a constituent quark. Yet, from the early SLAC days of the deep inelastic scattering on proton, we know that quark distribution in a proton shows precocious scaling for $Q^2$ in the range as low as one $GeV^2$. That is, $Q^2$ evolution has already run the course. For this reason, if $Q^2$ is small enough we may identify valon structure,$F^{valon}(z,Q^2)$ as $\delta(z-1)$ at some point, because we cannot resolve its internal structure at that $Q^2$ value.


\end{document}